\documentclass[epj,referee]{svjour}
\usepackage{graphics}
\begin{document}
\title{ Phase evolution of layered cobalt oxides versus
varying corrugation of the cobalt-oxygen basal plane}
\author{Hua Wu
\thanks{\emph{Present address:} Fritz-Haber-Institut der
Max-Planck-Gesellschaft, Faradayweg 4-6, D-14195 Berlin, Germany}%
}                     
%
%
\institute{Max-Planck-Institut f\"ur Physik komplexer Systeme, 
N\"{o}thnitzer Str. 38, D-01187 Dresden, Germany}
%
%
\abstract{
A general spin-state model and a qualitative physical
 picture
have been proposed for
a class of lately synthesized layered cobalt oxides
(LCOs) by means of density functional calculations.
As the plane corrugation of the cobalt-oxygen layer
decreases,
the LCOs evolve from
a high-spin (HS) superexchange-coupled antiferromagnetic (AFM)
insulator to
an almost-HS AFM/ferromagnetic (FM) competing
system where the FM coupling is mediated via the $p$-$d$
exchange by an increasing
amount of delocalized $pd\sigma$ holes having mainly the
planar O $2p$ character.
It is tentatively suggested
that the delocalized holes more
than 0.3 per CoO$_{2}$ basal square are likely necessary for the
insulator-metal and/or AFM-FM transitions in the
corrugation-weakened LCOs.
A phase control may be realized in LCOs by varying the plane
corrugation (thus modifying the hole concentration) through an
ionic-size change of the neighboring layers
on both sides of the cobalt-oxygen layer. In addition, a few
experiments are suggested for a check of the present model and
picture.
} 
\maketitle
\section{Introduction}
\label{1}
The phase diagrams of transition-metal oxides (TMOs)
concerning their crystal structures, electronic/magnetic
and transport properties drew very extensive attention in past
decades, especially after the discoveries of the
high-temperature superconducting
cuprates and the colossal magnetoresistance (CMR) manganites.
A great number of
experimental and theoretical studies have exploited rich
physics [1], which
not only leads one to understanding (although imcomplete so far) of the
nature and origin of their fascinating properties but also provides
a guide in search of novel function materials. As members of the $3d$
TMOs, the cobalt oxides are of current considerable interest.
This is, on one hand, due to their potential technological utilities such as
the analogs of the CMR perovskite manganite [2] $R_{1-x}A_{x}$MnO$_{3}$
($R$=rare earth; $A$=alkaline earth)---$R_{1-x}A_{x}$CoO$_{3}$ [3] and
$R$BaCo$_{2}$O$_{5+x}$ [4], and
the lately synthesized layered cobalt oxides (LCOs, analogs
of the superconducting cuprates) having a rich variety of phases
although no superconductivity observed in them up to now
---Sr$_{n+1}$Co$_{n}$O$_{2n+1}$Cl$_{n}$ (SCOC) 
are insulating with decreasing resistivity as the number of CoO$_{2}$
layers increases [5]; Sr$_{2}$Y$_{0.8}$Ca$_{0.2}$Co$_{2}$O$_{6}$ is an 
AFM insulator [6,7] while
Sr$_{2}$Y$_{0.5}$Ca$_{0.5}$Co$_{2}$O$_{7}$ is a magnetically glassy 
semiconductor and exhibits a pronounced FM interaction [6,8]; 
Bi$_{2}A_{3}$Co$_{2}$O$_{8+x}$ ($A$=Ca,Sr) are
semiconducting while the Ba analog has a metal to semiconductor transition [9]. 
On the other hand, this is in fundamental viewpoints mainly because of 
an intriguing but (in many cases) controversial issue [10-18]
on the Co spin state which plays a vital role
in the physical properties of cobalt oxides.
The Co spin state relies primarily on a competition between
the crystal field (CF) and Hund exchange coupling [17,18],
and it is also affected by a $pd$ covalent
effect [10-12], as well as
external factors, $e.$$g.$, varying temperature induces a complex
spin state transition as observed in LaCoO$_{3}$ [11].
In addition, a few relevant issues
concentrate on their electronic/magnetic structures and the closely
related transport behaviors.

  In this paper, by means of density functional theory
(DFT) [19]
calculations,
both a general spin-state model and a qualitative physical
picture have been set up for LCOs which have
a common structural feature---a plane corrugation of the
cobalt-oxygen layer. The DFT calculations show at least a
tendency that
as the plane corrugation decreases, the cobalt evolves from
a high-spin (HS) state to an almost HS
state with an increasing amount of delocalized $pd\sigma$ holes
having mainly the planar O $2p$ character. The former
is responsible for a superexchange (SE) [20] coupled
AFM insulating nature, while the latter
gives rise to a coexistence of
(and a competition between) the inherent AFM insulating behavior
and the hole-mediated $p$-$d$ exchange [21]
FM metallicity.
This picture, although seeming qualitative, consistently
accounts for the
rich and varying phases
of LCOs as seen below, and it leads to a suggestion of a possible phase
control mechanism in LCOs.
 
\section{Computational Details}
\label{2}
The above LCOs [5-9] each contain a basic structure block, that is
the deformed CoO$_{5}$ pyramid
where the Co ion moves out of the O$_{4}$ basal square and
into the pyramid. The $ab$-planar CoO$_{2}$ layer is formed by
those corner-shared CoO$_{4}$ ``squares" extended along the $a$
and $b$ axes. Thus a common plane corrugation actually exists in
the so-called CoO$_{2}$ layer. A parameter $D$ is defined here
for a description of the plane corrugation, and it represents
an average distance between the Co ion and the O$_{4}$ basal
``square".

  The DFT calculations have been performed for the
monolayered (1L) Sr$_{2}$CoO$_{3}$Cl and for the
bilayered (2L) Sr$_{3}$Co$_{2}$O$_{5}$Cl$_{2}$ and
Sr$_{2}$Y$_{0.8}$Ca$_{0.2}$Co$_{2}$O$_{6}$, since their
structure data are available among the above
LCOs [5,7].
Sr$_{2}$CoO$_{3}$Cl takes a tetragonal strucutre
($a$=3.9026 \AA~, $c$=14.3089 \AA~) at room-temperature (RT),
and so is 
Sr$_{3}$Co$_{2}$O$_{5}$Cl$_{2}$
($a$=3.9142 \AA~, $c$=24.0098 \AA~), while
Sr$_{2}$Y$_{0.8}$Ca$_{0.2}$Co$_{2}$O$_{6}$ takes an orthorhombic
structure ($a$=3.8291 \AA~, $b$=3.8238 \AA~, $c$=19.5585 \AA~)
at 260 K. Their $D$ parameters are 0.325, 0.285, and 0.44 \AA~,
respectively.

 Adopted in the DFT calculations is the full-potential
linearly combined
atomic orbitals band method [22], where no shape
approximation
is made for charge densities and potentials. A
full-potential method should be
preferable for descriptions of the lattice-distorted
materials, compared with an atomic-spherical approximation.
Sr $4p4d5s$, Co $3d4s$, O $2s2p$, and Cl $3s3p$ orbitals are
treated as valence states.
A virtual-crystal approximation is made for
Sr$_{2}$Y$_{0.8}$Ca$_{0.2}$Co$_{2}$O$_{6}$, where a virtual
atom with
nuclear charge $Z$=0.8$\times$$Z_{Y}$+0.2$\times$$Z_{Sr}$=38.8
takes the place
of the Y$_{0.8}$Ca$_{0.2}$ site. This virtual-crystal
approximation is acceptable, since in most cases the
Y [Ca(Sr)] atom only donates its
$4d^{1}5s^{2}$ [$4s^{2}$($5s^{2}$)] valence electrons and scarcely contributes
to the usually concerned valence bands around Fermi level ($E_{F}$). Thus,
the virtual atom takes the formal +2.8 valence ($4p^{6}4d^{0}5s^{0}$) as
Y$_{0.8}^{3+}$Ca$_{0.2}^{2+}$ in the real material. Hartree potential is
expanded in terms of lattice harmonics up to $L$=6, and an exchange-correlation
potential of von Barth-Hedin type [23] is taken within the local-(spin-)
density approximation [L(S)DA]. 8$\times$8$\times$2
($k_{x}\times$$k_{y}\times$$k_{z}$), 8$\times$8$\times$1, and
6$\times$6$\times$2 special ${\bf k}$ points in 1/8 irreducible Brillouin zone
are respectively used for the self-consistent calculations
of Sr$_{2}$CoO$_{3}$Cl, Sr$_{3}$Co$_{2}$O$_{5}$Cl$_{2}$, and
Sr$_{2}$Y$_{0.8}$Ca$_{0.2}$Co$_{2}$O$_{6}$.
A small number
of the ${\bf k}$ points are taken along the $c$ axis
due to the large
$c$-axis constant and to the two-dimensional
$ab$-planar structure.

  Although L(S)DA can give some useful information
for TMOs, its descriptions for TMOs are often unsatisfactory
quantitatively and even erroreous qualitatively in some cases,
$e.$$g.$, concerning the insulating ground state of FeO
and CoO [24].
This is because TMOs are commonly classified as strongly
correlated electron systems, whereas the L(S)DA in the framework
of mean-field approximation underestimates the electron
correlation effects. As such, the electron correlation should
be included in a better way into electronic structure
calculations of TMOs. Along this direction, on-site Coulomb
correlation correction (so called LSDA+$U$)
method [24] proves rather
powerful. For this reason, LSDA+$U$ calculations are
performed with $U$=5 eV for the Co $3d$ electrons [25].
As seen below,
the present discussion is made mainly on the basis of the
LSDA+$U$ results which are of more concern in this work.

\section{Results}
\label{3}
\subsection{Sr$_{2}$CoO$_{3}$Cl}
\label{3.1}
A spin-restricted LDA calculation is first performed for the paramagnetic
(PM) state of Sr$_{2}$CoO$_{3}$Cl in order to separate the crystal-field (CF)
effect from the magnetic exchange splitting. It can be seen in Fig. 1(a) that
the Co $3d$ bands lie between --1 and 2 eV relative to $E_{F}$, 
the $ab$-planar O $2p$ bands range approximately from --7.5 
to --2 eV, the apical
$c$-O $2p$ bands are relatively narrow and located between --5.5
to --2 eV,
and the apical Cl $3p$ bands are a little lower than the $c$-O $2p$ ones.
In contrast, the very narrow Sr $4p$, O $2s$, and Cl $3s$ bands lie at deep 
levels of --15, --18 eV, and that between them, respectively; whereas the 
empty Sr $4d5s$ bands lie above 5 eV and the Co $4s$ one even higher. Both the 
deep 
and high-level bands have a negligible contribution to the most concerned $pd$
valence bands
ranging from --8 to 2 eV, and therefore they are not shown here.   
In the RT tetragonal structure, the Co $3d$ $xz$/$yz$ doublet 
has the 
lowest level, which is followed by a litter higher $xy$ singlet.
The $x^{2}$--
$y^{2}$ level is highest and the $3z^{2}$--$r^{2}$ one is second highest. 
The narrow $xz$/$yz$ and $xy$ bands have weak $pd\pi$ 
hybridizations with the 
neighboring O $2p$ orbitals, as is indicated by their small component in the 
lower-level bonding states.  
While the $x^{2}$--$y^{2}$ and $3z^{2}$--$r^{2}$ bands have a large 
bonding-antibonding splitting due to strong $pd\sigma$ hybridizations, and
a pronounced bonding state between the $x^{2}$--$y^{2}$ and
the $ab$-O $p_{x,y}$ orbitals appears at --5.5 eV, and that between 
the $3z^{2}$--$r^{2}$ and $c$-O $p_{z}$ orbitals at --4.5 eV.
Whereas the Co $3d$ orbitals almost have no interaction with 
the Cl $3p$ one due to a long
Co-Cl distance of 3.116 \AA~, as is indicated by no 
corresponding bonding and antibonding states. In the sense, 
the CoO$_{5}$Cl octahedron in Sr$_{2}$CoO$_{3}$Cl can be 
viewed as a CoO$_{5}$ pyramid. 
In addition, a point-charge model calculation for the given 
structure of the
Co$^{3+}$O$^{2-}_{5}$Cl$^{-}$ octahedron shows that the 
Co $3d$ CF levels
are ordered in unit of eV as ($xz$/$yz$, 0, reference zero 
energy), ($3z^{2}$--$r^{2}$, 0.11),
($xy$, 0.23), and ($x^{2}$--$y^{2}$, 0.51).
A reverse of the $3z^{2}$--$r^{2}$ and $xy$ levels
in the above band data is attributed to the modified Coulomb potential arising
from a real charge distribution other than  
the idealized point
charge. It is evident that the Co$^{3+}$ ion sees a reduced CF 
in 
Sr$_{2}$CoO$_{3}$Cl due to a strong deformation of the CoO$_{5}$
pyramid, especially a strong plane corrugation 
$D$=0.325 \AA~. As a result, 
the Co$^{3+}$ ion strongly tends to take a high-spin (HS) ground state 
to lower exchange energy by Hund coupling [17]. 

  For the above reason, a spin-polarized LSDA calculation
is performed below for the assumed ferromagnetic (FM) state. The Co $3d$ and 
O $2p$ bands undergo significant changes due to strong magnetic exchange 
interactions. The majority-spin $t_{2g}$ ($xz$/$yz$ and $xy$) and the 
$3z^{2}$--$r^{2}$ orbitals become full-filled, and the 
majority-spin
$x^{2}$--$y^{2}$ orbital couples to the $ab$-O $p_{x,y}$ ones 
and hence forms
conduction bands wider than 3 eV, as seen in 
Fig. 1(b). 
The minority-spin $e_{g}$ ($3z^{2}$--$r^{2}$ and 
$x^{2}$--$y^{2}$) orbitals 
are nearly empty, while the minority-spin $t_{2g}$ bands, being near degenerate
and very narrow, just lie at $E_{F}$ and thus give a high DOS 
(density of states) peak at $E_{F}$. This DOS peak implies an 
instability of the RT tetragonal structure.
The calculated Co$^{3+}$ spin moment is 2.5 $\mu_{B}$, and the $pd$ 
hybridizations induce a non-negligible spin moment of 0.15 (0.31) $\mu_{B}$ at
the $ab$-O ($c$-O) site. The $ab$-O spin is smaller than the $c$-O one due to
a delocalization of the $ab$-O $p_{x,y}$ electrons. Whereas the Sr$^{2+}$ and 
Cl$^{-}$ ions have a very minor moment of only about 0.01 $\mu_{B}$.
Note that either the Co$^{3+}$ spin alone or the total spin of 3.1 $\mu_{B}$ per formula
unit (fu) is larger than a spin-only value of 2 $\mu_{B}$ for an intermediate-spin
(IS, $t_{2g}^{5}e_{g}^{1}$, $S$=1) state which turns out to be unstable and converges to
the present spin-polarized state. Moreover, the total energy of this state is lower than
that of the above low-spin (LS, $t_{2g}^{6}e_{g}^{0}$, $S$=0) PM state by 
0.02 eV/fu, which is significantly increased
up to 0.10 eV/fu in the LSDA+$U$ calculations. These results suggest that
the Co$^{3+}$ ion takes a HS ($t_{2g}^{4}e_{g}^{2}$, 
$S$=2) ground state in 
Sr$_{2}$CoO$_{3}$Cl [26], as is further supported by the  
following LSDA+$U$ calculations.   
 
 It can be seen in Fig. 2 that the electron correlations
significantly enhance the Co orbital- and spin-polarization. The Co $3d$ 
orbitals become strongly localized, and the majority-spin 
$3d$ orbitals are completely occupied except for the nearly 
full-filled $x^{2}$--$y^{2}$ one [27].
The $ab$-planar O $2p$ orbitals coupled to this 
$x^{2}$--$y^{2}$ one contribute to only 0.07 $pd\sigma$ holes
per CoO$_{2}$ basal square in the FM state as seen in Fig. 2(a).
The Co spin moment (total one per formula unit)
is increased up to 3.15 (3.8) $\mu_{B}$, both of which are close to
the HS state. It is evident that the upper valence bands (lower conduction 
bands) arise from the $ab$-O $2p$ states (the Co $3d$ ones), thus leading to
a classification of Sr$_{2}$CoO$_{3}$Cl as a $p$-$d$ 
charge-transfer (CT) oxide [28] like an ordinary late 
$3d$ TMO. 
The formation of the above $pd\sigma$ holes having mainly 
the $ab$-O $2p$ character is closely related to 
such a $p$-$d$ CT nature. Although delocalized $pd\sigma$ holes 
tend to induce a FM metallicity conceptually via the $p$-$d$ 
exchange [21] due to the dominant O $2p$ hole character 
(rather than the double-exchange [29]), the small
amount of 
(only 0.07)
holes are insufficient to induce an actual FM metallic state
for Sr$_{2}$CoO$_{3}$Cl. Instead a strong superexchange (SE) AFM 
coupling exists between the HS Co$^{3+}$ ions [20], and 
the dominant AFM coupling strongly suppresses the bandwidths and
naturally yields an insulating ground state with a large CT gap of
1.28 eV [see Fig. 2(b)] which has lower total energy than the above
FM state by 0.48 eV/fu. The calculated spin
moment of the HS Co$^{3+}$ ion is reduced to 3.17 $\mu_{B}$ in this AFM
state due to finite $pd$ hybridizations. Thus, the in-plane exchange
integral is estimated to be $J_{ab}\approx$48 meV, if Heisenberg model
applys.

\subsection{ Sr$_{3}$Co$_{2}$O$_{5}$Cl$_{2}$}
\label{3.2}
Since LSDA+$U$ calculations are more concerned in this work as 
stated in Sec. 2, here only the LSDA+$U$ results are shown.

 In the FM state (see Fig. 3), the majority-spin Co$^{3+}$ $t_{2g}$ and 
$3z^{2}$--$r^{2}$ orbitals are completely occupied, and the
majority-spin $x^{2}$--$y^{2}$ orbital is nearly full-filled.
While the little more half-filled minority-spin $xz$/$yz$ doublet,  
by allowing orbital polarization,  
splits and forms an $xz$/$yz$ orbital ordered (OO) state which is to be
stabilized by a cooperative Jahn-Teller (JT) distortion as discussed below.
In this CT-type LCO, the $ab$-planar O $2p$ orbitals couple
to the $x^{2}$--$y^{2}$ one and form wide conduction bands.
In particular, the majority-spin $pd\sigma$ band contributes to
0.22 delocalized holes per CoO$_{2}$ basal square. 
The increasing amount of $pd\sigma$ holes from 0.07 in 1L
Sr$_{2}$CoO$_{3}$Cl to 0.22 in 2L 
Sr$_{3}$Co$_{2}$O$_{5}$Cl$_{2}$ suggest a stronger electron
delocalization in the latter. Correspondingly, the calculated Co 
$3d$ ($ab$-O $2p$) spin moment decreases from 3.15 (0.17) 
$\mu_{B}$ in Sr$_{2}$CoO$_{3}$Cl to 3.10 (0.14) $\mu_{B}$
in Sr$_{3}$Co$_{2}$O$_{5}$Cl$_{2}$ as seen in Table 1. 
As seen in Ref. 5, Sr$_{2}$CoO$_{3}$Cl and
Sr$_{3}$Co$_{2}$O$_{5}$Cl$_{2}$ have an almost identical
$ab$-planar Co-O bond-length with a difference only 
0.001 \AA~. Naturally the enhanced electron 
delocalization in the latter can be ascribed to the 
decreasing $D$ from 0.325 \AA~ in the former to 0.285 \AA~
in the latter, which corresponds to an increasing $ab$-planar
Co-O-Co bond angle from 161.1 to 163.6 degree [5].

 The Co$^{3+}$ ion takes an almost HS state in the assumed
FM state of Sr$_{3}$Co$_{2}$O$_{5}$Cl$_{2}$ as 
shown above [26]. The 0.22 delocalized $pd\sigma$ holes could 
yield a FM metallic-like signal via the $p$-$d$ exchange
coupling [21]. Here it is recalled that in the layered-type
double perovskite TbBaCo$_{2}$O$_{5.5}$ with a decreasing
$D$ ($D$=0.19 and 0.32 \AA~ in the deformed CoO$_{6}$ 
octahedron and CoO$_{5}$ pyramid in the 270-K phase, 
respectively) [16], the delocalized $pd\sigma$ holes 
within the 
$ab$-plane are remarkably increased 
up to 0.61 per formula unit and they contain 0.305 per 
CoO$_{2}$ basal square on average [18]. Moreover, 
there are 0.24
holes which are delocalized along the $c$-axis CoO chain
formed by the apex-shared CoO$_{6}$ octahedra [18].
As discussed in Ref. 18, 
the increasing amount of holes in the almost HS state of  
TbBaCo$_{2}$O$_{5.5}$ are responsible for an appearance of a 
FM state at 260-340 K with a sudden drop of resistivity,
and for a high-$T$ PM metallic state. Whereas the strong 
HS-coupled AFM insulating behavior is inherent
and competes with the FM metallicity, thus leading to
FM-AFM and metal-insulator (M-I) transitions in 
TbBaCo$_{2}$O$_{5.5}$.
In the sense, a HS AFM insulating ground state is also
expected for 2L Sr$_{3}$Co$_{2}$O$_{5}$Cl$_{2}$, although 
a finite FM signal could appear due to the hole delocalization
with increasing $T$. However, no real calculation is attempted
here for the AFM state of Sr$_{3}$Co$_{2}$O$_{5}$Cl$_{2}$
because of a high computational cost for the large unit cell. 

\subsection{Sr$_{2}$Y$_{0.8}$Ca$_{0.2}$Co$_{2}$O$_{6}$}
\label{3.3}
LSDA+$U$ calculation for the assumed FM state of 
Sr$_{2}$Y$_{0.8}$Ca$_{0.2}$Co$_{2}$O$_{6}$ shows that the 
majority-spin $t_{2g}$ and $3z^{2}$--$r^{2}$ orbitals and 
the minority-spin $xy$ orbital are completely occupied. 
The majority-spin $x^{2}$--$y^{2}$ orbital couples to the
$a$- and $b$-axis O $2p$ ones and contributes to 0.13 $pd\sigma$
holes per CoO$_{2}$ square, while the minority-spin $xz$ narrow
band is partly occupied and just lies at $E_{F}$. As indicated
by experiments [7], $D$=0.44 \AA~ in 
Sr$_{2}$Y$_{0.8}$Ca$_{0.2}$Co$_{2}$O$_{6}$
is much larger than $D$=0.325 \AA~ in 
Sr$_{2}$CoO$_{3}$Cl [5].
A weaker $pd\sigma$ hybridization and thus a stronger electron
localization, and therefore a stronger ionicity should occur
in the former, as is supported by the present results (i)-(iv)
as seen in Table 1.
(i) The formal Co$^{2.6+}$ (Co$^{3+}$) ion takes the $3d^{6.4}$
($3d^{6}$) state in Sr$_{2}$Y$_{0.8}$Ca$_{0.2}$Co$_{2}$O$_{6}$
(Sr$_{2}$CoO$_{3}$Cl), while the calculated occupation 
number of the Co$^{2.6+}$ $3d$ orbital is about 0.1 smaller
than that of the Co$^{3+}$ one.
(ii) The calculated Co spin moment of 3.06 $\mu_{B}$ attains
85\% spin polarization of the HS Co$^{2.6+}$ (3.6 $\mu_{B}$) ion,
while the corresponding spin polarization is weaker than
80\% in both Sr$_{2}$CoO$_{3}$Cl and 
Sr$_{3}$Co$_{2}$O$_{5}$Cl$_{2}$. 
(iii) The $ab$-O and $c$-O ionic charge increases 
by 0.1-0.25 and 0.1, respectively. 
(iv) Reduced $pd$ 
hybridization induces a smaller oxygen spin moment, 0.04 
$\mu_{B}$ for the $a$- and $b$-axis oxygens, and 0.15 $\mu_{B}$ for 
the $c$-axis oxygen.
Owing to only a minor difference 
(smaller than 0.03 \AA~) [5,7]
between the $ab$-planar Co-O bond length of 
Sr$_{2}$Y$_{0.8}$Ca$_{0.2}$Co$_{2}$O$_{6}$ and that
of Sr$_{2}$CoO$_{3}$Cl, instead the variance of the $D$ 
parameter,
and the corresponding variance of the planar Co-O-Co 
bond angle from 161.1 degree in Sr$_{2}$CoO$_{3}$Cl [5] 
to 151.2/157.2 degree in 
Sr$_{2}$Y$_{0.8}$Ca$_{0.2}$Co$_{2}$O$_{6}$ [7]
are believed to play a leading role in their 
electronic-structure variances. Speaking qualitatively,
increasing $D$ would lead to a decreasing amount of planar
delocalized $pd\sigma$ holes in LCOs. Note that effective charge of
different Co valent state in LCOs affects the level 
distributions also. It is therefore not surprising
that the $pd\sigma$ hole number is larger in 
Sr$_{2}$Y$_{0.8}$Ca$_{0.2}$Co$_{2}$O$_{6}$ than in
Sr$_{2}$CoO$_{3}$Cl, because the Co$^{2.6+}$ $3d$ levels
of the former, lying higher than the Co$^{3+}$ ones of 
the latter, couple to the $ab$-O $2p$
states and contribute to a little more holes.    
However, these holes are more localized in 
Sr$_{2}$Y$_{0.8}$Ca$_{0.2}$Co$_{2}$O$_{6}$ than in
Sr$_{2}$CoO$_{3}$Cl as shown above.
 
  Owing to the large $D$ and strong hole 
localization in Sr$_{2}$Y$_{0.8}$Ca$_{0.2}$Co$_{2}$O$_{6}$,
this LCO is expected to be a HS AFM insulator like layered
$Ln$BaCo$_{2}$O$_{5}$ ($Ln$=Y,Tb,Dy,Ho) with 
$D\sim$0.4 \AA~ at RT [14,15,17],
although no real calculation is performed for the AFM state with 
a large unit cell.

\section{Discussion}
\label{4}
  The RT tetragonal structure of both Sr$_{2}$CoO$_{3}$Cl and
Sr$_{3}$Co$_{2}$O$_{5}$Cl$_{2}$ seems unstable because of the Co$^{3+}$
$t_{2g}$-level near degeneracy and/or the half-filled $xz$/$yz$ doublet, 
as indicated in the above
calculations. This is also a case for the layered 
$Ln$BaCo$_{2}$O$_{5}$ ($Ln$=Y,Tb,Dy,Ho) (see Fig. 2 in Ref. 17).
As observed experimentally [14,15], $Ln$BaCo$_{2}$O$_{5}$ 
actually 
undergo a Co$^{2+}$/Co$^{3+}$ charge-ordering (CO) transition 
and transform from the RT $Pmmm$ structure to the low-$T$ 
$Pmma$ one. A corresponding JT lattice distortion occurs in 
$Ln$BaCo$_{2}$O$_{5}$ and the resulting 
difference between the $a$- and $b$-axis Co-O bond lengths  
is about 0.05-0.1 \AA~ [15].
As a result, the near degeneracy 
of the $t_{2g}$ states is lifted in $Ln$BaCo$_{2}$O$_{5}$, and 
the minority-spin $xz$/$yz$ and $xy$ 
orbitals are occupied out of the $t_{2g}$ ones for the HS 
Co$^{2+}$ and Co$^{3+}$ ions, respectively [17]. 
Correspondingly, 
Sr$_{2}$CoO$_{3}$Cl and Sr$_{3}$Co$_{2}$O$_{5}$Cl$_{2}$ are
expected to transit into a low-$T$ distorted phase (likely 
orthorhombic) with alternate
$ab$-planar Co-O bond-lengths. A difference between these 
bond-lengths, about 0.05-0.1 \AA~ typical of a weak JT 
distortion seen by the $t_{2g}$ electrons, is believed to be 
large enough to lift the 
(near) degeneracy of the $t_{2g}$ orbitals and to strongly split the $xz$/$yz$ 
doublet with an aid of the $3d$ electron correlations [18],
thus giving 
rise to an $xz$/$yz$ OO insulating ground state for 
Sr$_{2}$CoO$_{3}$Cl and Sr$_{3}$Co$_{2}$O$_{5}$Cl$_{2}$.
If this OO state were a true ground state, a superlattice
diffraction would be observed. Moreover, it is believed that 
the reduced CF itself undergoes only a slight change in the OO 
state (as in the CO state of $Ln$BaCo$_{2}$O$_{5}$ [17]) 
and does not modify the above (almost) HS state or affect
the present discussion.

  One could also expect a presence of a finite Co orbital moment and a 
spin-orbital coupling (SOC) in Sr$_{2}$CoO$_{3}$Cl and
Sr$_{3}$Co$_{2}$O$_{5}$Cl$_{2}$, since the localized 
Co $3d$ electrons see
a reduced CF. However, the SOC commonly has an order of 
magnitude 
of only 0.01 eV in the $3d$ TMOs, and it is significantly weaker than the Hund 
coupling typical of $\sim$ 1 eV. Therefore, a negligence of the weak SOC
effect in the present calculations does not at least affect the present 
conclusion concerning the Co$^{3+}$ (almost) HS state.
But note that since  
the $t_{2g}$ orbitals are near degenerate in the RT tetragonal 
structure, 
the SOC could be operative by lowering the $L_{z}$=$\pm$1 
($xz\pm$${\bf i}$$yz$) singlets
at the alternate AFM-coupled Co sites, thus inducing alternate orbital 
moments of $\pm$1 $\mu_{B}$. 
In the sense, either this state or the above OO state calls for
a further study.
 
  It is evident that the lately synthesized LCOs belong to
two-dimensional systems in the viewpoints of both their 
crystallographical and electronic structures, and the  
plane corrugation of
the CoO$_{2}$ basal layer is a common structure feature of 
them, which leads to a reduced CF seen by the Co ions.
In the cobalt-trivalent SCOC series (1L Sr$_{2}$CoO$_{3}$Cl, 
2L Sr$_{3}$Co$_{2}$O$_{5}$Cl$_{2}$ and 3L 
Sr$_{4}$Co$_{3}$O$_{7.5}$Cl$_{2}$), the $ab$-planar
Co-O bond-lengths vary very slightly around 1.97 \AA~
within a range of 0.01 \AA~ [5]. 
While it was also found [5] that  
$D$ decreases monotonously
($D_{1L}$=0.325 \AA~$>$$D_{2L}$=0.285 \AA~
$>$$D_{3L}$=0.266 \AA~) in the SCOC series 
with the increasing number of the layers (due to
a covalance contraction caused by an interlayer coupling).
It is therefore firmly believed that the varying $D$ (and thus
varying planar Co-O-Co bond-angles) is essentially responsible 
for the evolutions of electronic, magnetic, and transport 
properties of the SCOC series. The decrease of $D$ in the 
SCOC series, as shown in the above calculations, leads to the 
rising $x^{2}$--$y^{2}$ level and 
enhanced planar $pd\sigma$ hybridizations and electron 
delocalization, and hence to an increasing amount of 
delocalized $pd\sigma$ 
holes in the almost HS state. The delocalized $pd\sigma$ holes
have mainly the $ab$-O $2p$ character in these CT-type 
SCOC. The mobile O $2p$ holes are antiferromagnetically coupled
to the Co $3d$ spins, and hence they tend to induce an 
effective FM coupling (between the Co spins) and a
metallic-like behavior via the $p$-$d$ exchange rather than
the double-exchange. However, the delocalized holes seem 
insufficient to induce an actual FM metallic state in SCOC,
compared with the case of the AFM-FM and I-M transition material
TbBaCo$_{2}$O$_{5.5}$ with smaller $D$ and more holes, as 
stated in Sec. 3.2. As indicated above,
Sr$_{2}$CoO$_{3}$Cl takes a HS AFM insulating ground state, and 
so is expected for Sr$_{3}$Co$_{2}$O$_{5}$Cl$_{2}$, and 
Sr$_{4}$Co$_{3}$O$_{7.5}$Cl$_{2}$ as well although no real 
calculation is performed for it due to a random distribution of
the 0.5 oxygen holes. 
While the increasing amount
of $pd\sigma$ holes consistently account for the decreasing
resistivities and effective magnetic moments ($\mu_{eff}$) 
as observed in SCOC [5] with the
increasing number of the layers.     
Since $D$ is larger in SCOC than in the layered-type
TbBaCo$_{2}$O$_{5.5}$ but smaller than in $Ln$BaCo$_{2}$O$_{5}$ 
as shown above, 
the electronic, magnetic, and transport properties of the
SCOC series can be qualitatively understood 
as an intermediate phase between the AFM-FM and I-M transition
material 
TbBaCo$_{2}$O$_{5.5}$ and the complete HS 
coupled 
SE-AFM insulator 
$Ln$BaCo$_{2}$O$_{5}$.

  The measured insulating behavior of 
the SCOC series [5] can be explained here. 
Whereas the original suggestion [5] of their magnetic 
properties differs much from the present results in that 
no magnetic ordering was detected and a small Co$^{3+}$ 
$\mu_{eff}$=0.80 (0.56) $\mu_{B}$ was extracted 
in the so-called LS Sr$_{2}$CoO$_{3}$Cl
(Sr$_{3}$Co$_{2}$O$_{5}$Cl$_{2}$). It seems that
the original suggestion was only based on a 
fitting of the inverse magnetic susceptibility ($\chi^{-1}$)
data measured below 250 K to a simple Cuire-Weiss 
law [5]. It is argued that 
such a measurement and a fitting are insufficient, since the AFM 
interaction temperature in SCOC ($e$.$g$., $J_{ab}\approx$48 meV
in Sr$_{2}$CoO$_{3}$Cl)
is expected to be near RT (probably between the 
$T_{\rm N}$=260 K 
of TbBaCo$_{2}$O$_{5.5}$ [16] and 330-340 K of 
$Ln$BaCo$_{2}$O$_{5}$ [14,15]). 
Therefore,
further experiments such as a measurement of $\chi$ 
above RT and a neutron diffraction would 
provide more informative magnetism data. Moreover, it is
worth noting that the present
conclusion on the HS state of Sr$_{2}$CoO$_{3}$Cl
is supported by a late joint study [30] of O-$K$ and 
Co-$L_{2,3}$ x-ray absorption spectra (XAS) and charge-transfer 
multiplet calculation. 

  Now we turn to the 2L 
Sr$_{2}$Y$_{1-x}$Ca$_{x}$Co$_{2}$O$_{6+\delta}$.
The structure of Sr$_{2}$Y$_{0.8}$Ca$_{0.2}$Co$_{2}$O$_{6}$ [7] 
can be viewed as a
SrY$_{0.8}$Ca$_{0.2}$Co$_{2}$O$_{5}$ bilayer plus a 
SrO monolayer. The 
SrY$_{0.8}$Ca$_{0.2}$Co$_{2}$O$_{5}$ bilayer containing the apical-oxygen
deficient CoO$_{5}$ pyramids shares a great similarity
to the structure of $Ln$BaCo$_{2}$O$_{5}$ ($Ln$=Y,Tb,Dy,Ho; 
$D\sim$0.4 \AA~ at RT) which are also 
of current interest [14,15]. 
The 0.13 holes per CoO$_{2}$ square in the assumed FM state
of Sr$_{2}$Y$_{0.8}$Ca$_{0.2}$Co$_{2}$O$_{6}$ are mainly due to
the rising Co$^{2.6+}$ $3d$ levels (compared with the 
Co$^{3+}$ case in above SCOC) and the corresponding 
$pd\sigma$ hybridizations. While these holes are more localized
than those in SCOC, because a larger $D$(=0.44 \AA~ in the 
260-K structure [7]) in 
Sr$_{2}$Y$_{0.8}$Ca$_{0.2}$Co$_{2}$O$_{6}$ strongly 
reduces the $ab$-planar $pd\sigma$ hybridizations. 
The present calculated Co spin moment of 
3.06 $\mu_{B}$ is close to an experimental value of 
2.93 $\mu_{B}$ [7] and 
attains 85\% spin polarization of the ideal 
3.6 $\mu_{B}$ for the formal HS Co$^{2.6+}$ ion, 
and it is also close to the previously calculated 3.1 $\mu_{B}$
for the HS Co$^{2.5+}$ (3.5 $\mu_{B}$) ion in the AFM 
$Pmmm$ structure of YBaCo$_{2}$O$_{5}$ [17].
It is therefore not surprising that the HS
Sr$_{2}$Y$_{0.8}$Ca$_{0.2}$Co$_{2}$O$_{6}$ 
(Sr$_{2}$Y$_{1-x}$Ca$_{x}$Co$^{2.5+}_{2}$O$_{6-\delta}$)
is an AFM insulator [6,7] with a high 
$T_{\rm N}\approx$ 270 K
(300 K) like $Ln$BaCo$_{2}$O$_{5}$ [14,15].
While both the formal non-integer Co$^{2.6+}$ valence and 
a partial filling
of the minority-spin $xz$ narrow band (but a full filling
of the lower-level $xy$ band due to the large $D$) imply an 
instability of 
Sr$_{2}$Y$_{0.8}$Ca$_{0.2}$Co$_{2}$O$_{6}$ 
against a structure distortion, which could induce a 
superstructure and 
form a charge disproportion and/or OO at low-$T$. Thus the holes 
become more strongly localized.     

  For Sr$_{2}$Y$_{1-x}$Ca$_{x}$Co$_{2}$O$_{6+\delta}$,
it can be assumed that the apical-oxygen holes of the
CoO$_{5}$ pyramids are gradually removed as $\delta$ increases, 
and instead more CoO$_{6}$ octahedra appear and thus $D$ 
decreases as clearly seen in TbBaCo$_{2}$O$_{5+\delta}$ 
($\delta$=0,$D\sim$0.4 \AA~; $\delta$=0.5,
averaged $D\sim$0.25 \AA~) [15,16].
Thus it is naturally expected, according to the above
calculations and discussion, that as $\delta$ increases 
($D$ decreases) the Co 
ion evolves from a complete HS state to an almost HS state with
an increasing amount of delocalized $pd\sigma$ holes.
Note that this expectation is also supported
by the late O-$K$ and Co-$L_{2,3}$ XAS study [30] 
for Sr$_{2}$Y$_{0.5}$Ca$_{0.5}$Co$_{2}$O$_{6+\delta}$
($\delta$=--0.248,0.196,0.645).  
The increasing amount of delocalized holes 
tend to induce (via the $p$-$d$ exchange) 
a FM metallic signal in large $\delta$
(and high cobalt valence) samples, which competes with
the inherent HS-coupled SE-AFM insulating behavior.
Thus a possible spin frustration
gives rise to a spin-glassy insulating (semiconducting) state 
for the Co$^{3.4+}$ (Co$^{3.75+}$) sample 
with decreasing resistivity compared with 
the Co$^{2.6+}$ one 
Sr$_{2}$Y$_{0.8}$Ca$_{0.2}$Co$_{2}$O$_{6}$ [6,8].

  As discussed above, the $D$ parameter (and thus planar Co-O-Co
bond angles) is essentially responsible for the physical 
properties of the lately synthesized LCOs, although their 
electronic structures are affected by both the effective
charge of different cobalt valences and the planar Co-O bond
lengths. The present LSDA+$U$ calculations indicate at least a
tendency that as $D$ decreases in the LCOs, the Co ion 
evolves from a HS state to an almost HS state with an increasing amount of 
the delocalized $pd\sigma$ holes having mainly the planar
O $2p$ character in these CT-type oxides. 
The former is responsible for a SE-coupled AFM
insulating nature, while the latter gives rise to a coexistence 
of (and a competition between) the inherent AFM insulating 
behavior and the
hole-mediated $p$-$d$ exchange FM metallicity.
It is an variance of the $pd\sigma$ hole number that
modifies the electronic, magnetic, and transport properties
of the LCOs. Thus a general spin-state model and a qualitative
physical picture have been proposed for the LCOs. 

  It was found in experiments that the SCOC series
($D\le$0.266 \AA~) are all insulating [5], and that
TbBaCo$_{2}$O$_{5.5}$ (an averaged $D\sim$0.25 \AA~)
exhibits the AFM-FM and I-M transitions [16].
The LSDA+$U$ calculations show that 
Sr$_{3}$Co$_{2}$O$_{5}$Cl$_{2}$ and TbBaCo$_{2}$O$_{5.5}$
have 0.22 and 0.305 planar delocalized holes per CoO$_{2}$
basal square, respectively. A combination of both the
experimental findings and the present results leads to
a tentative suggestion that more than 0.3 delocalized holes
per CoO$_{2}$ square are likely necessary for the  
I-M and/or AFM-FM transitions in LCOs, where $D$ is
probably smaller than 0.25 \AA~.
As such, the physical properties of LCOs can be modified by
hole doping ($D$ variance) through element substitutions,
particularly through an ionic-size change of the neighboring 
layers on both sides of the cobalt-oxygen layer. 

  A large ionic-size difference between the apical O$^{2-}$ and 
Cl$^{-}$ ions (1.40 \AA~ $vs$ 1.81 \AA~, 
six coordinated [31]) in the CoO$_{5}$Cl octahedron 
should be a 
structural origin for the plane corrugation
in the above SCOC series. It can be
postulated that F$^{-}$ doping or substitution for the Cl$^{-}$, if operative, 
could make $D$ smaller than 0.25 \AA~, since the F$^{-}$ 
radius [31] of 1.33 \AA~ is near the O$^{2-}$ size. In this way,  
the I-M and/or AFM-FM transitions could occur in the 
F$^{-}$ doped 
or substituted SCOC with increasing hole number. 
In particular, a cobalt analog of the fascinating layered 
manganite
(La,Sr)$_{n+1}$Mn$_{n}$O$_{3n+1}$ [32] could 
emerge with FM metallic layers. 
Apart from the parent 
insulator SCOC, a few likely intriguing properties
of the F$^{-}$ doped or substituted SCOC, like CMR of 
La$_{2-2x}$Sr$_{1+2x}$Mn$_{2}$O$_{7}$ [32] or 
superconductivity of 
the layered Sr$_{2}$CuO$_{2}$F$_{2+\delta}$ [33] may deserve  
experimental investigations.  

 Moreover, the present picture could be employed for the 2L 
Bi$_{2}A_{3}$Co$_{2}$O$_{8+\delta}$ ($A$=Ca,Sr,Ba).
It can be reasonably assumed that with increasing 
$A$ size, the planar oxygens
move towards the apical-oxygen-deficient $A$ layer 
(see Fig. 1 in Ref. 9) in a 
decreasing tendency, thus leading to a decreasing $D$.
Thus the present picture could account for 
qualitatively the observed tendency [9] that 
Bi$_{2}A_{3}$Co$_{2}$O$_{8+\delta}$ varies from 
the semiconducting
behavior for smaller $A$=Ca and Sr (with larger $D$) to the 
semiconductor-metal transition for larger $A$=Ba 
(with smaller $D$), as well as their increasing metallicity 
through lead doping (hole doping).  

  Furthermore, the present spin-state model may be  
an alternative even for the 1L La$_{2-x}$Sr$_{x}$CoO$_{4}$ 
containing an ideal CoO$_{2}$ plane ($D$=0), for which
an IS state model [12] and an ordered HS-LS model 
[13] were previously suggested separately. 
With an increasing Sr doping
(hole doping) in the well-defined HS AFM parent insulator 
La$_{2}$CoO$_{4}$, the 
$ab$-planar Co-O bond-length decreases monotonously from 
1.96 \AA~ ($x$=0) to 1.91 \AA~ ($x$=1) [12,34], which 
implies both an enhanced delocalization of the majority-spin 
$pd\sigma$ hybridized $x^{2}$--$y^{2}$ state and a gradual electron 
removal from (and thus hole injection into) the $pd\sigma$ 
antibonding state. While the large $c$-axis Co-O bond-length,
$e.$ $g.,$ for
$x$=0 and 1, varies slightly around 2.05 \AA~ [34], 
indicating that the localized majority-spin $3z^{2}$--$r^{2}$ 
orbital remains full-filled.
As a result, the increasing amount of delocalized $pd\sigma$
holes in the almost HS state, as in the above cases of SCOC,
naturally account for the consistently decreasing 
resistivity, $\mu_{eff}$ and Weiss temperature as 
observed [12] in La$_{2-x}$Sr$_{x}$CoO$_{4}$ upon 
Sr doping. Moreover, it was observed that  
their optical conductivity spectra ($E\parallel$$ab$)
evolve from a single O $2p$-Co $3d$ CT transition at $\sim$ 
3 eV for $x$=0.5 to two CT transitions at $\sim$ 2 eV and
$\sim$ 3 eV for $x$=1.0 [35]. 
According to the present almost HS model, the additional 
structure at $\sim$ 2 eV is ascribed to the transition
from the $ab$-O $2p$ state to the delocalized majority-spin 
Co $3d$ $x^{2}$--$y^{2}$ hole state.
If polarized ($E\parallel$$c$) spectra are to be measured, 
we suggest, the structure 
at $\sim$ 2 eV will disappear according to the present model, 
whereas it will remain according to the ordered HS-LS 
model [13] where the $3z^{2}$--$r^{2}$ hole state, 
besides the $x^{2}$--$y^{2}$ one, is also available.   
To some extent such a measurement can check 
an applicability of the present model to   
La$_{2-x}$Sr$_{x}$CoO$_{4}$. 

\section{Conclusions}
\label{5}
The lately synthesized LCOs---analogs of the superconducting 
layered cuprates and CMR layered manganites---exhibit rich 
and varying electronic, magnetic and transport properties [5-9]. 
They have a common structural feature---a plane corrugation ($D$) 
of the cobalt-oxygen layer, which gives rise to a reduced CF 
seen by the cobalt. As $D$ decreases, the cobalt 
evolves from a HS to an almost HS with an increasing 
amount of the delocalized $pd\sigma$ holes having mainly 
the planar O $2p$ character in these CT-type LCOs, as well 
evidenced by the present LSDA+$U$ calculations and supported
by a late O-$K$ and Co-$L_{2,3}$ XAS study [30]. 
A competition
between the HS-coupled SE-AFM insulating behavior 
and the delocalized-hole mediated FM metallicity (via the 
$p$-$d$ exchange rather than the double-exchange) is 
responsible for a rich variety of phases of LCOs. 
The present picture, although seeming qualitative, is rather 
general for LCOs by a series of demonstrations.
It is tentatively suggested that
more than 0.3 delocalized holes per CoO$_{2}$ basal square
are likely necessary for the I-M and/or AFM-FM transitions
in LCOs with $D$ probably smaller than 0.25 \AA~.
A phase control may be realized in LCOs 
by varying $D$ (thus modifying the hole concentration) 
through an ionic-size change of the neighboring
layers on both sides of the cobalt-oxygen layer, which leads 
to a search for likely novel LCOs.
It could be also somewhat meaningful for other layered TMOs.  
Moreover, the present spin-state model may be  
applicable to those cobalt-based systems containing a 
basal-square-corrugated CoO$_{5}$ 
pyramid or CoO$_{6}$ octahedron. In addition, a few
measurements of crystallography, magnetism, and spectroscopy
are suggested for a check of the present model and picture.\\

{\noindent \small The author is grateful to Z. Hu $et$ $al.$ [29] 
for communication of their results prior to 
publication and discussion with him, and thanks P. Fulde and 
A.-m. Hu for their 
suggestions. The author acknowledges Max-Planck Scholarship and hospitality of
MPI-PKS.}\\
\newpage
\begin{table} 
\caption{Electron occupation number/spin moment
(in $\mu_{B}$) calculated by LSDA+$U$ ($U$=5 eV) for the FM 
and/or AFM states.}
\label{Table 1}
\begin{tabular} {lccccccc}
 && \multicolumn{2}{c}{Sr$_{2}$CoO$_{3}$Cl} && Sr$_{3}$Co$_{2}$O$_{5}$Cl$_{2}$ && Sr$_{2}$Y$_{0.8}$Ca$_{0.2}$Co$_{2}$O$_{6}$ \\
 && FM & AFM && FM && FM \\ \hline
Co $3d$&&7.13/3.15&7.12/3.17&&7.09/3.10&&7.02/3.06\\
$a$-O $2p$&&5.32/0.17&5.30/0&&5.30/0.14&&5.43/0.04\\
$b$-O $2p$&&5.32/0.17&5.30/0&&5.30/0.14&&5.56/0.04\\
$c$-O $2p$&&5.24/0.32&5.23/0&&5.19/0.26&&5.32/0.15\\
$c$-Cl $3p$&&5.73/0.03&5.72/0&&5.75/0.02&&\\

\end{tabular}
\end{table}
\begin{figure*}
\vspace{10cm}       
\resizebox{1.0\textwidth}{!}{%
  \includegraphics{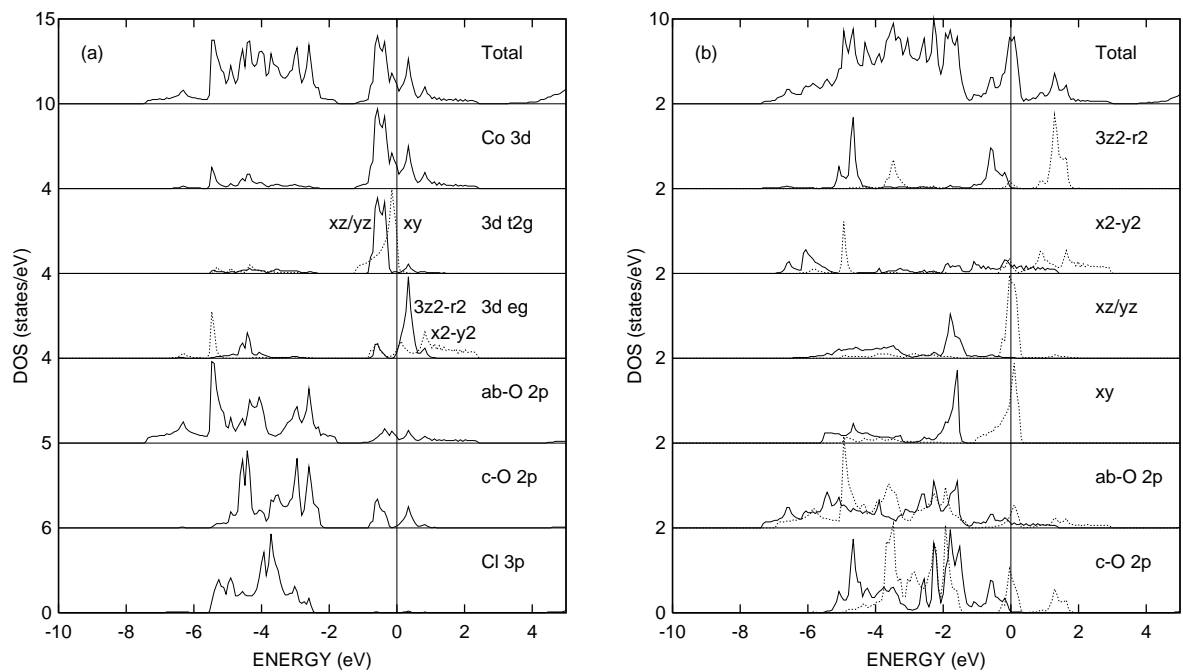}
}
\vspace{-10cm}       
\caption{ The (a) LDA-PM and (b) LSDA-FM DOS of
     Sr$_{2}$CoO$_{3}$Cl. Fermi level is set at zero.
     For the orbital-resolved DOS in (b), the solid (dashed)
     line denotes the majority (minority)
    spin channel. See discussions in main text.}
\label{Fig.1}       
\end{figure*}
\begin{figure*}
\vspace{12cm}       
\resizebox{0.9\textwidth}{!}{%
  \includegraphics{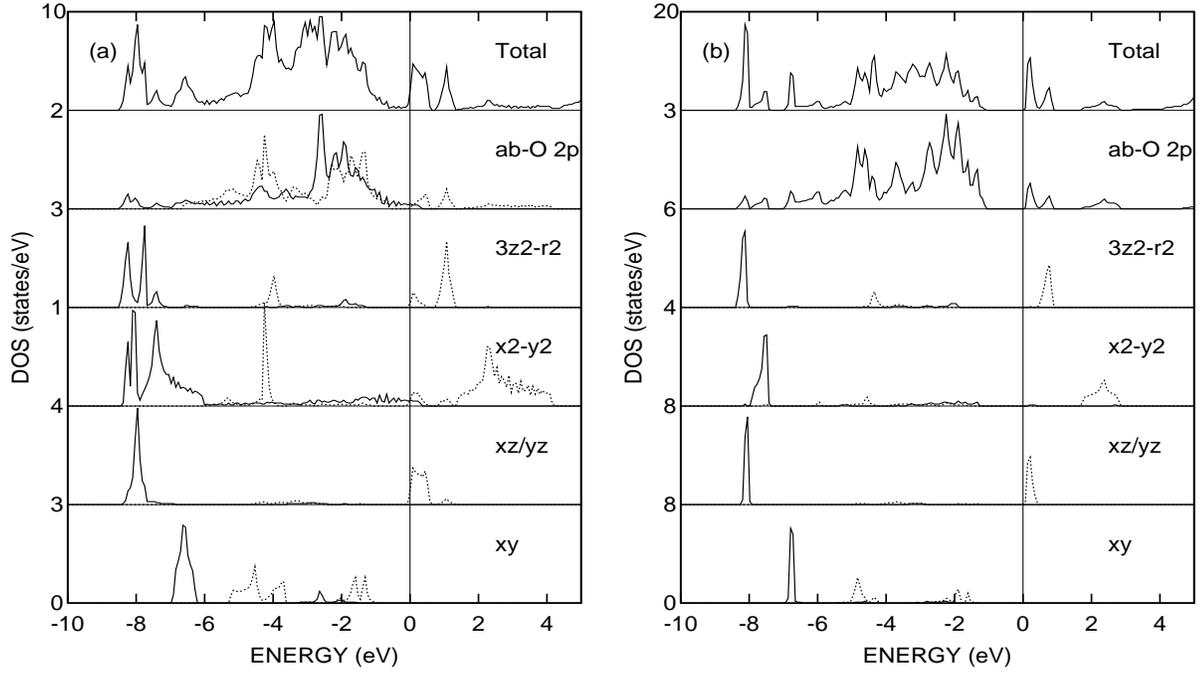}
} 
\vspace{-10cm}       
\caption{The LSDA+$U$ DOS for the (a) FM and (b)
     AFM states of Sr$_{2}$CoO$_{3}$Cl. (a) It is evident that
     the Co$^{3+}$ ion takes a HS state and only 0.07 $ab$-planar
     $pd\sigma$ holes per CoO$_{2}$ basal square have mainly the
     $ab$-O 2p character in this CT-type TMO.
     (b) The HS-coupled AFM insulating ground state is stabilized
     by opening a CT gap of 1.28 eV.}
\label{Fig.2}       
\end{figure*}
\begin{figure*}
\vspace{10cm}       
\resizebox{1.0\textwidth}{!}{%
  \includegraphics{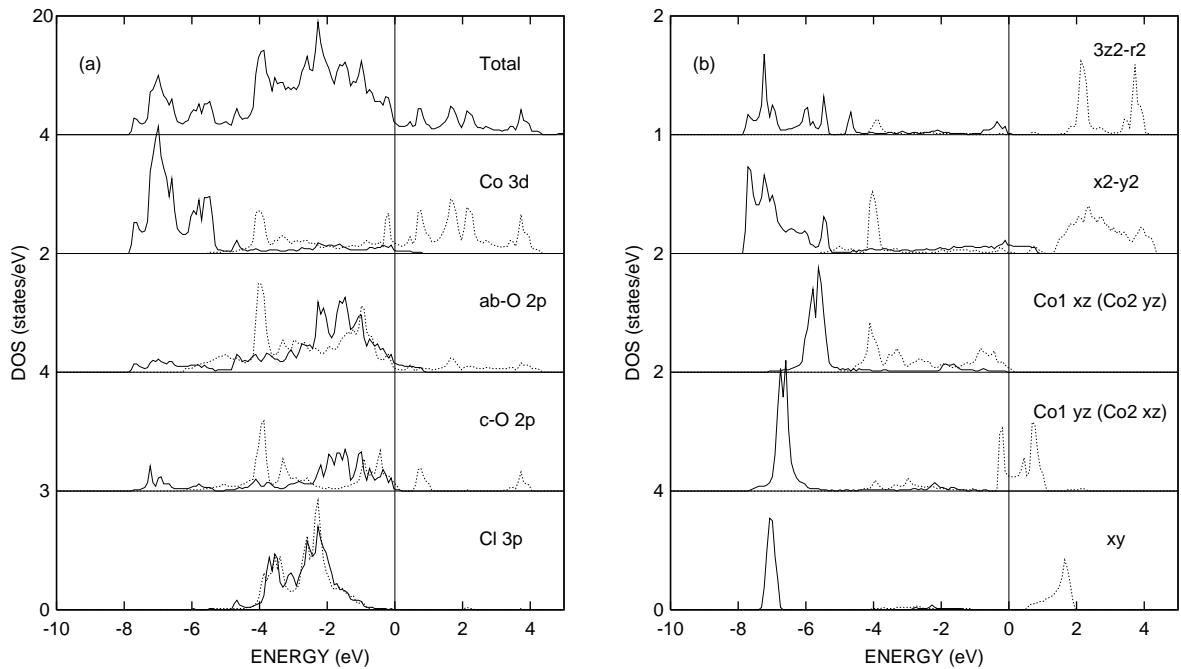}
}
\vspace{-10cm}       
\caption{The LSDA+$U$ DOS for the FM
     state of Sr$_{3}$Co$_{2}$O$_{5}$Cl$_{2}$.
     The solid (dashed) line denotes the majority (minority) spin.
     The Co$^{3+}$ ion takes an almost HS state, and there are
     0.22 $ab$-planar
     $pd\sigma$ holes per CoO$_{2}$ basal square.
     This FM state, with an $xz$/$yz$ OO to be stabilized by a JT
     lattice distortion, is expected to evolve into a HS-AFM insulating
     ground state as shown in Fig. 2.}
\label{Fig.3}       
\end{figure*}
\begin{figure*}
\vspace{10cm}       
\resizebox{1.0\textwidth}{!}{%
  \includegraphics{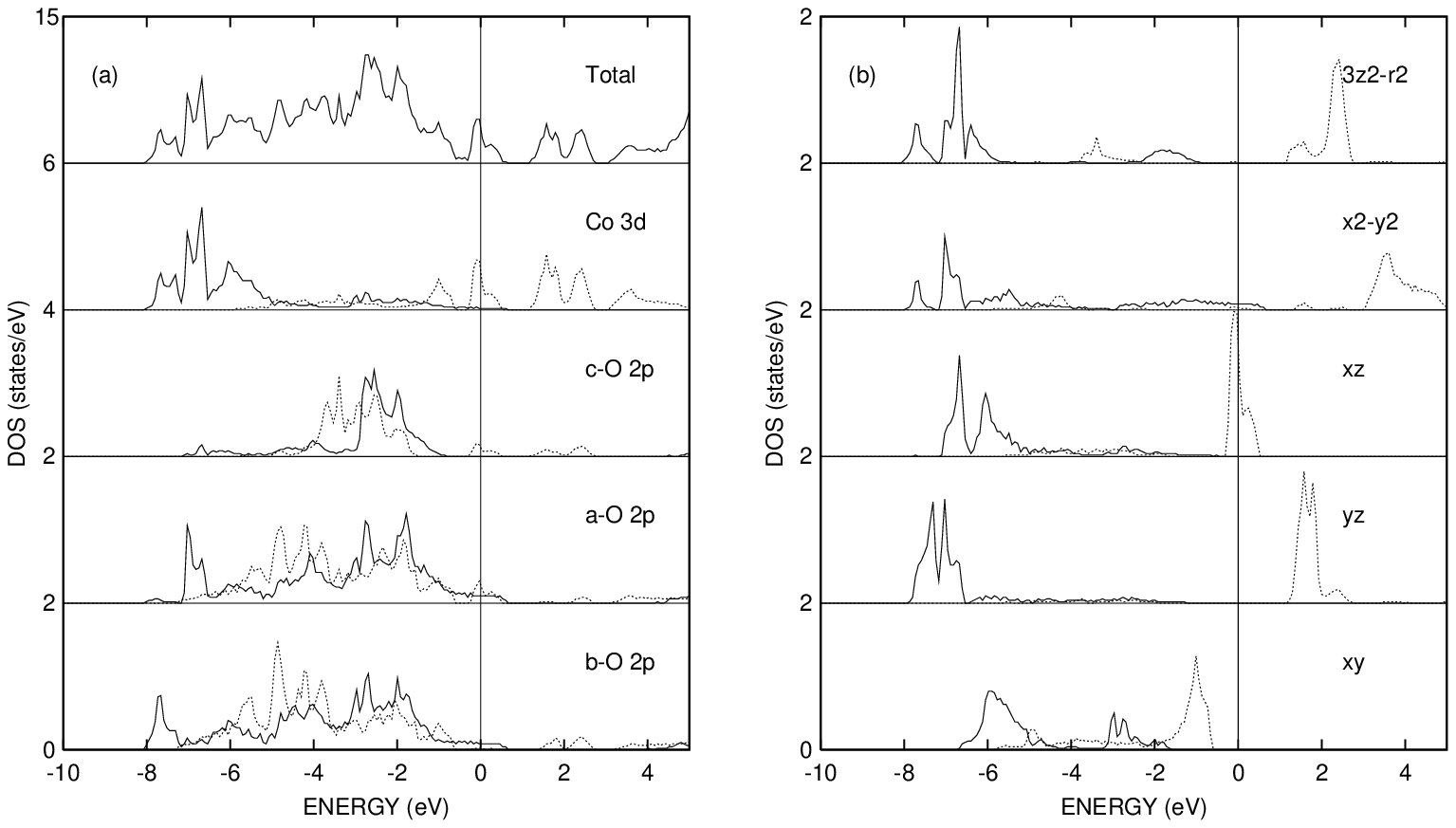}
}
\vspace{-10cm}       
\caption{ The LSDA+$U$ DOS for the FM
    state of Sr$_{2}$Y$_{0.8}$Ca$_{0.2}$Co$_{2}$O$_{6}$.
    The Co$^{2.6+}$ ion takes a HS state, and
    the 0.13 $ab$-planar
    $pd\sigma$ holes per CoO$_{2}$ basal square
    are mainly due to the rising Co$^{2.6+}$ $3d$ levels
    [for comparison see the Co$^{3+}$ $3d$ levels shown
    in Fig. 2(a)].}
\label{Fig.4}       
\end{figure*}

\begin{thebibliography}{}
\bibitem{c26}  Strongly in contrast, the octahedrally coordinated Co ion in
the pseudocubic perovskite LaCoO$_{3}$ sees a strong CF and takes 
a LS ground state [17], while a strong $pd\sigma$ hybridization
stabilizes an IS state as the $t_{2g}$ valence electrons are thermally
excited to the broad $e_{g}$ conduction bands with increasing temperature [11].
However, the pyramidally coordinated Co ion sees a weak CF in the LCOs and
the Co $3d$ electrons have a decreasing $pd\sigma$ hybridization both due to
the apical-oxygen deficiency and the strong base corrugation. 
The mechanism of forming the LS (IS) state by a strong CF (a strong $pd\sigma$
hybridization) seems ineffective in the LCOs, and therefore the HS state is 
preferable instead.
\bibitem{c27}  There seems to be an uncertainty concerning the $t_{2g}$ 
levels in Sr$_{2}$CoO$_{3}$Cl due to their near degeneracy. Both the LDA 
calculation [Fig. 1(a)] and the point-charge model suggest that 
the $xz$/$yz$ level is lower than the $xy$ one as in 
Sr$_{3}$Co$_{2}$O$_{5}$Cl$_{2}$
[Fig. 3(b)]. Whereas the
minority-spin $xy$ orbital has a little larger occupation 
number ($e.$$g.$, $n$=0.57 in the LSDA-FM calculation) 
than the $xz$/$yz$ doublet ($n$=0.51$\times$2) because of a 
stronger $pd$ hybridization for the former.
Consequently, the $xy$ level is 
lowered and becomes full-filled (Fig. 2) by the $U$ correction 
with the expression like
$V^{LSDA+U}_{m\sigma}$=$V^{LSDA}_{\sigma}$+
($\frac{1}{2}$--$n_{m\sigma}$)$U_{eff}$. 
This uncertainty is to be removed provided that the $t_{2g}$ near
degeneracy is lifted even by a weak Jahn-Teller distortion as discussed in 
the text.
\end{thebibliography}
\end{document}